\documentclass[sigconf]{acmart}

\usepackage{comment}
\usepackage{graphicx}
\usepackage{balance}  
\usepackage{amsmath}
\usepackage{bm}
\usepackage{tabularx}

\usepackage[caption=false]{subfig}

\usepackage{multirow}
\usepackage{mathtools}
\usepackage[mathscr]{eucal}
\usepackage{natbib}

\usepackage[linesnumbered,ruled,algonl,vlined,noend]{algorithm2e}
\usepackage{amsmath}
\usepackage{color}
\setcitestyle{usenames,dvipsnames}
\usepackage{slashbox}

\usepackage{enumitem}
\usepackage{booktabs}
\usepackage[T1]{fontenc}
\usepackage[utf8]{inputenc}
\usepackage{xcolor}

\SetKwRepeat{Do}{do}{while}

\usepackage{amsthm} 

\theoremstyle{definition}
\newtheorem{definition}{Definition}

\usepackage{algorithmic}

\let\vec\mathbf

\renewcommand{\algorithmiccomment}[1]{\bgroup\hfill//~#1\egroup}

\usepackage[normalem]{ulem}
\usepackage{soul}
\usepackage{xcolor}

\usepackage{url}

\newcommand{\mycaption}[1]{\caption{\normalfont{#1}}}

\newcommand{\myparagraph}[1]{\vspace{0.1\baselineskip}\noindent{\textbf{#1.}}~}

\newcommand{\heatmapscale}{0.152}

\newcommand{\interactionSet}{\ensuremath{\mathcal{I}}\xspace}

\newcommand{\matrixP}{\ensuremath{\vec{P}^l}\xspace}
\newcommand{\matrixQ}{\ensuremath{\vec{Q}^l}\xspace}

\newcommand{\scorefinal}{\ensuremath{\vec{O}^l}\xspace}
\newcommand{\predScoreInter}{\ensuremath{\vec{S}^l}\xspace}
\newcommand{\predScoreHis}{\ensuremath{\vec{G}^l}\xspace}

\newcommand{\modelnameshort}{\ensuremath{\mathsf{MPR}}\xspace}
\newcommand{\modelnamelong}{multi-level POI recommendation\xspace}
\newcommand{\taskOne}{attribute-based representation learning}
\newcommand{\taskTwo}{interaction-based representation learning}

\newcommand{\scoreOne}{\ensuremath{s_{ij}}\xspace}

\newcommand{\explainFeature}{user-aspect\xspace}
\newcommand{\explainPOI}{POI-aspect\xspace}
\newcommand{\explainHistory}{interaction-aspect\xspace}

\newcommand{\checkinSeq}{\ensuremath{\mathcal{Q}}\xspace}
\newcommand{\checkinSeqSubset}{\ensuremath{\mathcal{Q}_s}\xspace}

\newcommand{\adjustPara}{\ensuremath{\tau}\xspace}

\newcommand{\expParaHis}{\ensuremath{\gamma}\xspace}

\newcommand{\poiTreeLayerOne}{\ensuremath{H_1}\xspace}
\newcommand{\poiTreeLayerTwo}{\ensuremath{H_2}\xspace}
\newcommand{\poiTreeLayerThree}{\ensuremath{H_3}\xspace}

\newcommand{\poiTree}{\ensuremath{\mathcal{T}}\xspace}
\newcommand{\poiTreeNum}{\ensuremath{L}\xspace}

\newcommand{\poiSet}{\ensuremath{\mathcal{P}}\xspace}
\newcommand{\poiNum}{\ensuremath{n_l}\xspace}
\newcommand{\poiOne}{\ensuremath{p_j}\xspace}
\newcommand{\poiPos}{\ensuremath{p_j^{+}}\xspace}
\newcommand{\poiNeg}{\ensuremath{p_j^{-}}\xspace}

\newcommand{\featureNum}{\ensuremath{f}\xspace}
\newcommand{\featureNumUser}{\ensuremath{f_u}\xspace}
\newcommand{\featureNumPoi}{\ensuremath{f_p}\xspace}

\newcommand{\poiAttTimes}{\ensuremath{tu_{ijk}}\xspace}
\newcommand{\poiAttTimesMax}{\ensuremath{tu_{jk}^{\uparrow}}\xspace}
\newcommand{\poiAttTimesMin}{\ensuremath{tu_{jk}^{\downarrow}}\xspace}

\newcommand{\poiTreeNode}{\ensuremath{p_\poiTreeIndex^\poiTreeLayer}\xspace}
\newcommand{\poiTreeNodeChildren}{\ensuremath{p_j^{l+1}}\xspace}
\newcommand{\poiTreeNodeParent}{\ensuremath{p_i^l}\xspace}

\newcommand{\poiTreeLayerChildren}{\ensuremath{H_{l}}\xspace}

\newcommand{\poiTreeLayer}{\ensuremath{l}\xspace}
\newcommand{\poiTreeIndex}{\ensuremath{i}\xspace}

\newcommand{\poiAttMatrix}{\ensuremath{\vec{YA}^l}\xspace}
\newcommand{\poiTagMatrix}{\ensuremath{\vec{YT}^l}\xspace}

\newcommand{\userSet}{\ensuremath{\mathcal{U}}\xspace}
\newcommand{\userNum}{\ensuremath{m}\xspace}
\newcommand{\userOne}{\ensuremath{u_i}\xspace}

\newcommand{\userAttMatrix}{\ensuremath{\vec{XA}}\xspace}
\newcommand{\userTagMatrix}{\ensuremath{\vec{XT}}\xspace}

\newcommand{\userTagTimes}{\ensuremath{tp_{ijk}}\xspace}
\newcommand{\userTagTimesMax}{\ensuremath{tp_{ik}^{\uparrow}}\xspace}
\newcommand{\userTagTimesMin}{\ensuremath{tp_{ik}^{\downarrow}}\xspace}

\newcommand{\poiAttentionWeight}{\ensuremath{\vec{W}^{l+1}}\xspace}
\newcommand{\poiAttentionHidden}{\ensuremath{d_1}\xspace}

\newcommand{\poiImplictEmbeddingChild}{\ensuremath{\vec{h}_j^{l+1}}\xspace}

\newcommand{\poiAttImplicitEmbeddingParent}{\ensuremath{\vec{a}_i^l}\xspace}

\newcommand{\poiAttParent}{\ensuremath{p_i^l}\xspace}
\newcommand{\poiAttChild}{\ensuremath{p_j^{l+1}}\xspace}

\newcommand{\poiGeospatialGraph}{\ensuremath{\mathcal{G}}\xspace}
\newcommand{\poiGeospatialVertex}{\ensuremath{\mathcal{V}}\xspace}
\newcommand{\poiGeospatialEdge}{\ensuremath{\mathcal{E}}\xspace}
\newcommand{\poiGeospatialMatrix}{\ensuremath{\vec{U}_g^l}\xspace}

\newcommand{\poiCoQuery}{\ensuremath{\delta(p_i, p_j|\Delta t_1)}\xspace}
\newcommand{\poiCoVisit}{\ensuremath{\psi(p_i, p_j|\Delta t_2)}\xspace}
\newcommand{\poiCoDist}{\ensuremath{\zeta(p_i, p_j)}\xspace}

\newcommand{\poiExampleCoQuery}{\ensuremath{\delta(\poiExampleOne, \poiExampleTwo|\Delta t_1)}\xspace}
\newcommand{\poiExampleCoVisit}{\ensuremath{\psi(\poiExampleOne, \poiExampleTwo|\Delta t_2)}\xspace}
\newcommand{\poiExampleCoDist}{\ensuremath{\zeta(\poiExampleOne, \poiExampleTwo)}\xspace}
\newcommand{\poiExampleOne}{\ensuremath{p_1}\xspace}
\newcommand{\poiExampleTwo}{\ensuremath{p_2}\xspace}

\newcommand{\userFeatureMatrix}{\ensuremath{\vec{X}}\xspace}
\newcommand{\userWithHiddenMatrix}{\ensuremath{\vec{U}_u}\xspace}
\newcommand{\userHiddenMatrix}{\ensuremath{\vec{H}_u^l}\xspace}
\newcommand{\poiFeatureMatrix}{\ensuremath{\vec{Y}^l}\xspace}
\newcommand{\poiWithHiddenMatrix}{\ensuremath{\vec{U}_p^l}\xspace}
\newcommand{\poiHiddenMatrix}{\ensuremath{\vec{H}_p^l}\xspace}

\newcommand{\poiLevelMatrix}{\ensuremath{\vec{A}_p^l}\xspace}
\newcommand{\userLevelMatrix}{\ensuremath{\vec{A}_u^l}\xspace}

\newcommand{\featurelatent}{\ensuremath{\vec{V}^T}\xspace}

\newcommand{\coefficientAttribute}{\ensuremath{\lambda_1}\xspace}
\newcommand{\coefficientInteraction}{\ensuremath{\lambda_2}\xspace}

\newcommand{\lossTotal}{\ensuremath{\mathcal{L}}\xspace}
\newcommand{\lossAttribute}{\ensuremath{\mathcal{L}_1}\xspace}
\newcommand{\lossInteraction}{\ensuremath{\mathcal{L}_2}\xspace}

\newcommand{\tup}[0]{\ensuremath{^\triangle}\xspace}
\newcommand{\btup}[0]{\ensuremath{^\blacktriangle}\xspace}
\newcommand{\tdown}[0]{\ensuremath{^\triangledown}\xspace}
\newcommand{\btdown}[0]{\ensuremath{^\blacktriangledown}\xspace}

\AtBeginDocument{%
	\providecommand\BibTeX{{%
			\normalfont B\kern-0.5em{\scshape i\kern-0.25em b}\kern-0.8em\TeX}}}

\copyrightyear{2020} 
\acmYear{2020} 
\setcopyright{acmlicensed}\acmConference[SIGIR '20]{Proceedings of the 43rd International ACM SIGIR Conference on Research and Development in Information Retrieval}{July 25--30, 2020}{Virtual Event, China}
\acmBooktitle{Proceedings of the 43rd International ACM SIGIR Conference on Research and Development in Information Retrieval (SIGIR '20), July 25--30, 2020, Virtual Event, China}
\acmPrice{15.00}
\acmDOI{10.1145/3397271.3401090}
\acmISBN{978-1-4503-8016-4/20/07}

\begin{document}
\fancyhead{}
\title{Spatial Object Recommendation with Hints: When Spatial Granularity Matters}

\author{Hui Luo{$\scriptstyle ^{1,\dagger}$}, Jingbo Zhou{$\scriptstyle ^{*,2,3}$}, Zhifeng Bao{$\scriptstyle ^{1}$}, Shuangli Li{$\scriptstyle ^{2,4}$}, }
\author{J. Shane Culpepper{$\scriptstyle ^{1}$}, Haochao Ying{$\scriptstyle ^{5}$}, Hao Liu{$\scriptstyle ^{2,3}$}, Hui Xiong{$\scriptstyle^{*,6}$}}
\thanks{$\dagger$This work was done when Hui Luo visited Baidu Research.}
\thanks{$^*$Jingbo Zhou and Hui Xiong are corresponding authors.}
\affiliation{$^{1}$RMIT University, $^{2}$Business Intelligence Lab, Baidu Research}
\affiliation{$^{3}$National Engineering Laboratory of Deep Learning Technology and Application, China}
\affiliation{$^{4}$University of Science and Technology of China, $^{5}$Zhejiang University, $^{6}$Rutgers University}
\email{{hui.luo,zhifeng.bao,shane.culpepper}@rmit.edu.au }
\email{{zhoujingbo, liuhao30, v\_lishuangli}@baidu.com, haochaoying@zju.edu.cn,hxiong@rutgers.edu}

\def\authors{Hui Luo, Jingbo Zhou, Zhifeng Bao, Shuangli Li, J. Shane Culpepper, Haochao Ying, Hao Liu, Hui Xiong}
\renewcommand{\shortauthors}{Luo, et al.}









\begin{abstract} \label{sec:abstract}
Existing spatial object recommendation algorithms generally treat
objects identically when ranking them.
However, spatial objects often cover different levels of spatial
granularity and thereby are heterogeneous.
For example, one user may prefer to be recommended a region (say
{\textit{Manhattan}}), while another user might prefer a venue (say a
{\textit{restaurant}}).
Even for the same user, preferences can change at different stages of
data exploration.
In this paper, we study how to support top-$k$ spatial object
recommendations at varying levels of spatial granularity, enabling
spatial objects at varying granularity, such as a city, suburb, or
building, as a Point of Interest (POI).
To solve this problem, we propose the use of a {\em POI tree}, which
captures spatial containment relationships between POIs.
We design a novel multi-task learning model called {\modelnameshort}
(short for \textit{\textbf{M}ulti-level \textbf{P}OI
\textbf{R}ecommendation}), where each task aims to return the top-$k$
POIs at a certain spatial granularity level.
Each task consists of two subtasks: (i) \taskOne; (ii)
\taskTwo.
The first subtask learns the feature representations for both users
and POIs, capturing attributes directly from their profiles.
The second subtask incorporates user-POI interactions into the model.
Additionally, \modelnameshort can provide insights into why certain
recommendations are being made to a user based on three types of
hints: \textit{\explainFeature},
\textit{\explainPOI}, and \textit{\explainHistory}.
We empirically validate our approach using two real-life datasets,
and show promising performance improvements over several
state-of-the-art methods.
\end{abstract}


\maketitle

\vspace{-2ex}
\section{Introduction} \label{sec:introduction}
Spatial object recommendation is an important location-based service
with many practical applications, where the most relevant venues
\cite{ye2011exploiting,yuan2014graph} or regions
\cite{pham2017general} are recommended based on spatial, temporal,
and textual information.
Existing spatial object recommendation methods
\cite{ma2018point, yang2017bridging,wang2018exploiting} usually do
not differentiate the granularity of spatial objects (i.e., building
versus suburb), when ranking a list of top-$k$ objects.
However, the most appropriate granularity of spatial object ranking
may vary at different stages of data exploration for a user, and can
vary from one user to another, which is hard to predict a priori.
Choosing the most appropriate spatial granularity based on the
recommendation scenario is often
critical~\cite{bao2015recommendations}.
For example, if a user is planning to visit America for a holiday,
they may initially want to be recommended a particular region such as
``Los Angeles'' or ``New York'' at the beginning of data exploration.
The user might also wish to drill down for specific venue
recommendations such as a restaurant or a bar as the exploration
continues.

Therefore, user expectations at varying spatial granularity of POIs
(Point of Interests, i.e., a region or a venue) should be satisfied
by the recommender system adaptively and dynamically.
Note that a recommended region or venue is referred to as a POI for
ease of readability in this paper.
We refer to this as the {\textit{\modelnamelong}} problem, which aims
to recommend the top-$k$ POI candidates from each level of spatial
granularity.
Dynamic selection of the most appropriate recommendation level(s) is
driven by user interactions and application constraints.
Elucidating all integration-specific details of our proposed model
for an end-to-end production system is beyond the scope of this
paper.

To solve this {\textit{\modelnamelong}} problem, a straightforward
solution is to build a separate recommendation model for each level
of spatial granularity, and then apply an existing POI recommendation
algorithm directly.
However, this approach has one drawback: it may not fully leverage
mutual information among POIs at different spatial granularity
levels.
For example, a user may prefer to visit an area because of the POIs
contained in that area.
Therefore, a major challenge must be addressed: {\textit{How can we achieve a one-size-fits-all model to
make effective recommendations at every level of spatial
granularity?}}
In other words, instead of designing a best-match recommendation
model for level independently, the spatial containment relationships
between all of the levels should be considered.
If a user has visited a POI $p$ (like a shop), there will be a
check-in record for the parent POI(s) (like a shopping mall) covering
$p$.
Such information can heavily influence and assist the recommendation
of the parent POI(s).
\begin{figure}[t]
	\centering
	\includegraphics[width=7.8cm]{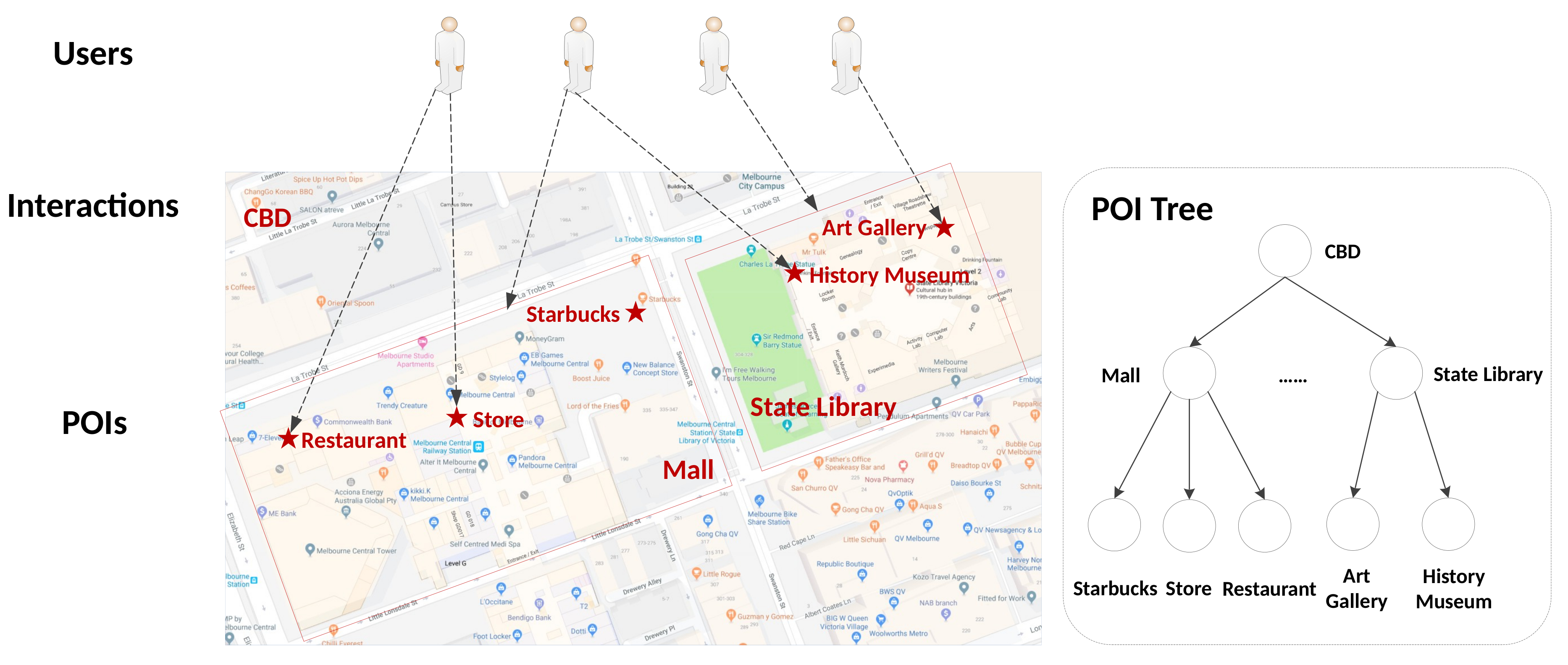}
	\vspace{-2ex}
	\mycaption{User check-in records at varying spatial
		granularity.}
	\label{fig-motivation}
	\vspace{-2em}
\end{figure}

In this paper, POIs are structured as a tree based on their spatial
containment --- defined as the relationship of a child POI which
is fully covered by a higher level POI~\cite{kalay1982determining}.
For example, a restaurant is within a mall, which in turn is within a
suburb (CBD) of a city in Figure~\ref{fig-motivation}, allowing
recommendation to be made at any level (i.e., a particular spatial
granularity) in the POI tree.
We then propose a new technique called \modelnameshort (short for
\textbf{M}ulti-level \textbf{P}OI \textbf{R}ecommendation), which
employs multi-task learning in order to jointly train the model
using every available level of spatial granularity.
Each task corresponds to recommending POIs located at a certain
spatial granularity.
Our approach is able to leverage data that is much sparser than prior
work~{\cite{li2016point,liu2014general,wang2018exploiting,yang2017bridging}},
which used only the check-in metadata found in commonly datasets such
as Foursquare or Gowalla.
Our two test collections were generated using real-life data from an
online map service which is also more heterogeneous than the
collections commonly used in similar studies.
Moreover, the sparsity of a user-POI check-in matrix for Foursquare
and Gowalla -- the most commonly used ones by existing work -- is
around $99.9\%$~\cite{liu2017experimental}, while our datasets are
much more sparse (i.e., around $99.97\%$), sparse, which is an
essential hurdle to overcome when using real map data.
In order to alleviate the sparsity issue, POI features can be
propagated from bottom to top in a POI tree using an attention
network mechanism, such that the information of child POI(s) can be
used by a parent POI in recommendations.
In essence, child POIs are learned features that contribute directly
to any related higher-level POIs, and multiple levels of such a
parent-child relationship can exist.
In addition, it is non-trivial to consider the geospatial influence
of a location when ranking a recommendation
\cite{liu2017experimental}.
That is, users are more likely to prefer nearby locations over
distant ones when they have a choice.
Thus we create a \textit{POI context graph} to describe the
geospatial influence factors between any two POIs at the same level,
which maps three different sources of spatial relationships ---
\textit{co-search}, \textit{co-visit}, and \textit{geospatial
distance}.

Lastly, it is worth noting that our proposed model can be used to
directly justify recommendations to a user for any level of spatial
granularity.
Providing justification for recommendations has been shown to be an
important factor in user satisfaction~\cite{baral2018reel,baral2017pers}.
For instance, when Alice is in a dilemma about a recommended POI, we
can provide recommendation hints along with the recommended POIs, in
three aspects where the latter two are unique in our model.
We can provide: (1) \textit{\explainFeature} hint based on the user 
profile: Chinatown appears to be an important
area since she loves dumplings based on her user profile.
(2) \textit{\explainPOI} hint based on the POI tree:
the particular region (such as the CBD) is initially recommended
since it contains several relevant shops and restaurants of interest
to the user.
(3) \textit{\explainHistory} hint based on the POI context graph: the
\textit{State Library} is also recommended because she has visited a
library several times before.

In summary, we make the following contributions:
\begin{itemize}[leftmargin=*]
\item We are the first to explore the multi-level POI recommendation
problem, which aims to simultaneously recommend POIs at different
levels of spatial granularity (Section~\ref{sec:pf}).
\item We propose a novel model \modelnameshort using multi-task
learning, where each task caters for one level of the spatial
granularity.
Each task has two subtasks: \textit{\taskOne}
(Section~\ref{sec:feature}) and \textit{\taskTwo}
(Section~\ref{sec:interaction}).
\item Our model can provide specific hints on why certain POI
recommendations are being made, namely \textit{\explainFeature},
\textit{\explainPOI}, and \textit{\explainHistory} hints
(Section~\ref{sec:explain}).
\item We perform extensive experiments on two large-scale real-life datasets to evaluate the performance of our
model.
Our experimental results show promising improvements over several state-of-the-art POI recommendation algorithms (Section \ref{sec:exp}).
\end{itemize}

\begin{table}[t]
	\scriptsize
	\renewcommand{\arraystretch}{1.3}
	\mycaption{A summary of key notations.
	The midlines partition the variables by section -- Section 2,
	3, and 4 respectively.
	}
	\vspace{-2ex}
	\label{table-symbol}
	\begin{tabular}{p{1.1cm}p{5.3cm}p{1.1cm}}
		\toprule
		Symbol	&	Description	&	Dimension	\\
		\midrule
		\userSet, \poiSet	&	The set of users and POIs, respectively	\\
		\interactionSet	&	The set of user-POI interactions	\\
		\poiTree	&	The POI tree with \poiTreeNum levels	\\
		$l$	&	The $l$-th level of \poiTree	\\
		\userNum (or \poiNum)	&	The number of users (or POIs at the $l$-th level of \poiTree)	\\ 
		\poiTreeNode	&	The \poiTreeIndex-th POI node located at the \poiTreeLayer-th level of \poiTree	\\
		$C(\poiTreeNodeParent)$	&	The child POIs rooted at \poiTreeNodeParent	\\ \bottomrule
		\featureNumUser (or \featureNumPoi)	& The number of users' (or POIs') explicit features; \featureNum=\featureNumUser+\featureNumPoi.	\\
		$r$	&	The latent factor size of explicit features.	\\
		\userFeatureMatrix (or \poiFeatureMatrix)	&	The observed matrix between users (or POIs at the $l$-th level of \poiTree) and their attributes	& $\mathbb{R}^{m \times \featureNum}$ (or $\mathbb{R}^{\poiNum \times \featureNum}$)	\\
		\userWithHiddenMatrix (or \poiWithHiddenMatrix)	&	The explicit feature representations of users (or POIs at the $l$-th level of \poiTree)	& $\mathbb{R}^{m \times r}$ (or $\mathbb{R}^{\poiNum \times r}$)	\\
		\featurelatent	&	The shared latent explicit feature representations of both users and POIs at the $l$-th level of \poiTree	& $\mathbb{R}^{r \times \featureNum}$	\\
		\userAttMatrix (or \poiAttMatrix)  &	The direct attribute matrix of users (or POIs at the $l$-th level of \poiTree) & $\mathbb{R}^{m \times \featureNumUser}$ (or $\mathbb{R}^{n_l \times \featureNumPoi}$)	\\
		\userTagMatrix (or \poiTagMatrix)  &	The inverse attribute matrix of users (or POIs at the $l$-th level of \poiTree) & $\mathbb{R}^{m \times \featureNumPoi}$ (or $\mathbb{R}^{n_l \times \featureNumUser}$)	\\	\bottomrule
		\poiPos, \poiNeg	&	The positive and negative POI instances \\
		$r_l$	&	The latent factor size of implicit features at the $l$-th level of \poiTree	\\
		$d_1$	&	The hidden layer size of the attention network	\\
		\scorefinal	&	The user-POI check-in matrix	& $\mathbb{R}^{m \times n_l}$	\\
		\predScoreInter  &	The feature-based check-in matrix	& $\mathbb{R}^{m \times n_l}$	\\
		\predScoreHis  &	The historical check-in matrix	& $\mathbb{R}^{m \times n_l}$	\\	
		$\vec{H}_u^{l}$ (or $\vec{H}_p^{l}$)	&	The implicit feature representations of users (or POIs at the $l$-th level of \poiTree)	& $\mathbb{R}^{m \times r_l}$ (or $\mathbb{R}^{n_l \times r_l}$)	\\
		\poiLevelMatrix	&	The inner-level propagated POI feature representation	&	$\mathbb{R}^{n_l \times r_{l+1}}$\\
		\bottomrule
	\end{tabular}
	\vspace{-4ex}
\end{table}

\vspace{-1em}
\section{Problem Formulation and Model Overview}\label{sec:pf}
Throughout this paper, all vectors are represented by bold lowercase
letters and are column vectors (e.g., $\vec{x}$), where the
$i$-th element is shown as a scalar (e.g., $\vec{x}_i$).
All matrices are denoted by bold upper case letters (e.g.,
$\vec{M}$); the element located in the $i$-th row and $j$-th column
of matrix $\vec{M}$ is marked as $\vec{M}_{i,j}$.
Also, we use calligraphic capital letters (e.g., \userSet) to denote
sets and use normal lowercase letters (e.g., $u$) to denote scalars.
Note that, the superscript $l$ is used in certain symbols to denote
the $l$-th level of \poiTree, such as \poiFeatureMatrix.
For clarity of exposition, Table~\ref{table-symbol} summarizes the
key notations used in this work, where only the dimensions of
matrices are reserved.
\vspace{-1ex}
\subsection{Problem Definition}
\begin{figure*}[t]
	\centering
	\includegraphics[width=14cm]{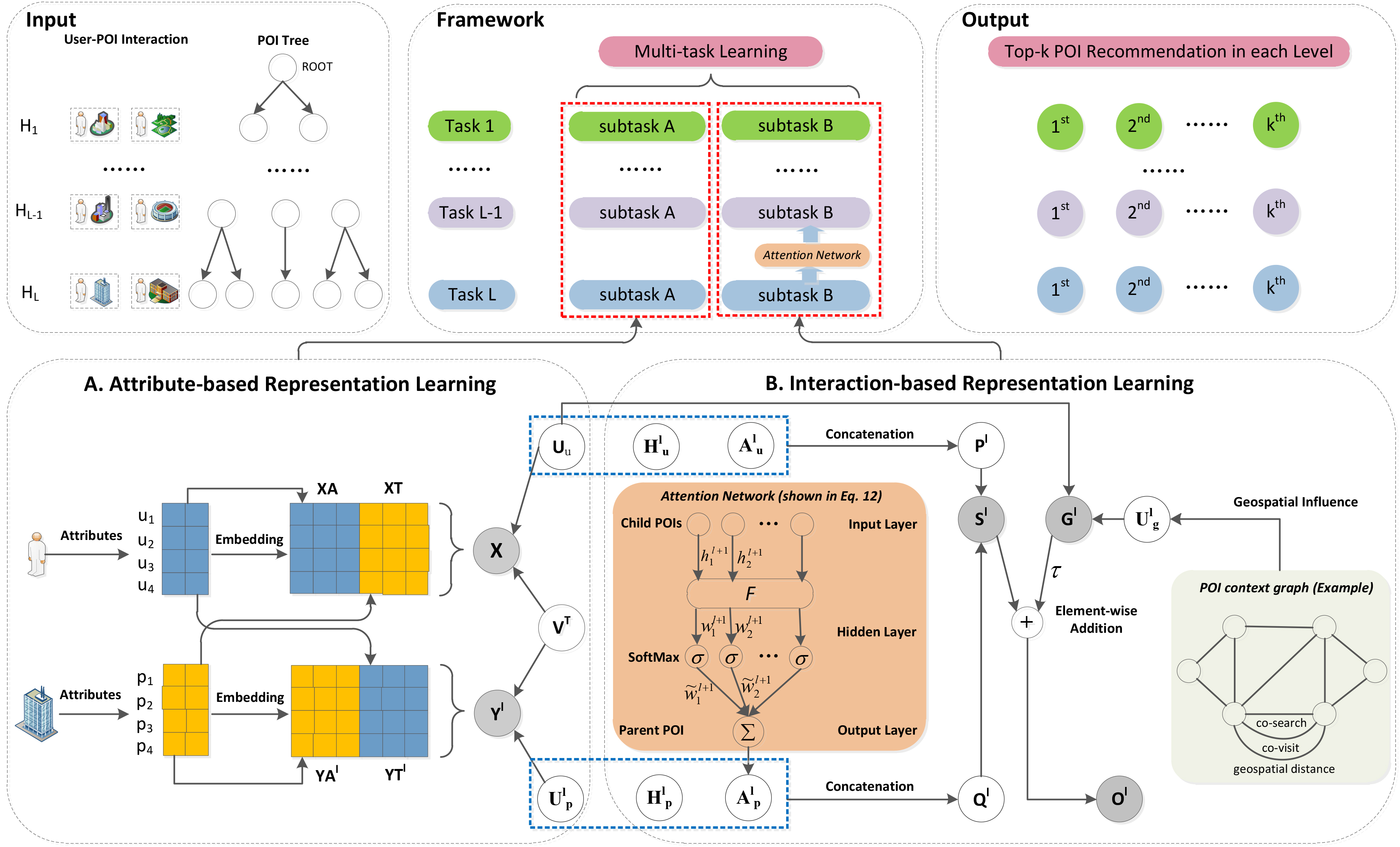}
	\vspace{-2ex}
	\mycaption{Architecture of the \modelnamelong (\modelnameshort)
		model. (Best viewed in color)}
	\label{fig-framework}
	\vspace{-3ex}
\end{figure*}
In a recommender system, there are a set of users \userSet $= \{u_1,
u_2, ..., u_m\}$ and a set of POIs \poiSet $= \{p_1, p_2, ..., p_n\}$
available.
Each user \userOne $\in$ \userSet has an attribute set derived from a
user profile, such as age and hobby.
Each POI \poiOne $\in$ \poiSet has two components: (i) a parent POI,
indicating that \poiOne is covered geospatially, and the parent POI
may be empty if \poiOne is a root area; (ii) an attribute set, which
is derived from the POI profile and typically contains attributes
such as a tag or category.
Based on spatial containment relationships among POIs, we construct a
POI tree (see Definition~\ref{def-poiTree}) over \poiSet to predict
POIs for each level of spatial granularity.
\begin{definition}
\label{def-poiTree}
(POI Tree) A POI tree \poiTree is a tree structure of \poiTreeNum
levels, where each node represents a spatial object,
\poiTreeLayerChildren denotes the $l$-th level of \poiTree, and
\poiNum is the number of POI nodes at level \poiTreeLayerChildren.
A node \poiTreeNodeParent is the parent of a node
\poiTreeNodeChildren if \poiTreeNodeParent contains
\poiTreeNodeChildren in geo-space.
We denote $C(\poiTreeNodeParent)$ as all child POIs rooted at
\poiTreeNodeParent. 
An illustrative example of a POI tree is shown in
Figure~\ref{fig-motivation}.
\end{definition}
\vspace{-1ex}
\myparagraph{User-POI Interaction}
Each instance of the interaction \interactionSet
between a user \userOne and a POI \poiOne is a tuple $\langle
\userOne, \poiOne, \scoreOne \rangle$, where the score \scoreOne
corresponds to a ``binary value'', indicating whether \userOne has
visited \poiOne (e.g., $\scoreOne = 1$ when \userOne has checked in
\poiOne; otherwise, $\scoreOne = 0$).

\begin{definition}
\label{def-problem}
(Multi-level POI Recommendation) Given a user, their historical
user-POI interactions, a pre-built POI tree \poiTree, and a parameter
$k$, return the top-$k$ most relevant POIs at each level of \poiTree.
\end{definition}

\vspace{-2ex}
\subsection{An Overview of the MPR Model}\label{subsec-approachOverview}
\myparagraph{Model Architecture}
The architecture of the model \modelnameshort is shown
in Figure~\ref{fig-framework}.
Taking the input of historical user-POI interactions and a pre-built
POI tree \poiTree based on spatial containment relationship,
{\modelnameshort} outputs the top-$k$ POIs for each level of
\poiTree.
To achieve the goal shown in Definition~\ref{def-problem}, we
leverage multi-task learning to implement a joint optimization over
all levels of the POI tree, where each task includes two main
subtasks for the given POI level:
\textit{\taskOne} (Section \ref{sec:feature}) and \textit{\taskTwo}
(Section \ref{sec:interaction}).

The first subtask explores the attributes of both users and POIs by
mapping them to two embedding spaces:
\userFeatureMatrix and \poiFeatureMatrix.
These are induced from two sources of information: (i) \userAttMatrix
and \poiAttMatrix, which are attributes directly derived from the
user and POI profile, respectively.
(ii) \userTagMatrix and \poiTagMatrix, which are derived from the
user and POI attribute distributions obtained from check-in
statistics.

The second subtask focuses on how to model the interactions between
users and POIs to further capture personal preferences.
Additionally, we model two important matrices: (i) the inter-level
POI features matrix \poiLevelMatrix propogated from child POIs
using an attention mechanism; and (ii) the geospatial
influence matrix \poiGeospatialMatrix between POIs derived from a
\textit{POI context graph} (Section~\ref{subsubsec-history_checkin}),
each edge of which contains one of the three types of spatial
relationships between any two POIs at the same level, i.e.,
\textit{co-search}, \textit{co-visit}, and \textit{geospatial
distance}.

These two subtasks are combined using shared latent factors (i.e.,
\userWithHiddenMatrix and \poiWithHiddenMatrix), in order to
guarantee that the feature representations of users and POIs at the
$l$-th level of \poiTree will remain unchanged despite attributes and
interactions being modeled in separate subtasks.

\myparagraph{Objective}
As each task in \modelnameshort incorporates two different learning
objectives for each subtask, we train a joint model by optimizing the
sum of two loss functions as follows.
\begin{equation}
\label{equ-lossShort}
\lossTotal = \coefficientAttribute \lossAttribute + \coefficientInteraction \lossInteraction + ||\Theta||_F^2
\vspace{-.5ex}
\end{equation}
where \lossAttribute and \lossInteraction are the loss functions for
the first and second objectives applied across all levels of \poiTree.
The computational details of these two loss functions are described
further in Section~\ref{sec:feature} and
Section~\ref{sec:interaction}, respectively.
\coefficientAttribute and \coefficientInteraction are hyper-parameters
to balance the trade-off between the two loss functions, and
$||\Theta||_F^2$ is the L2 regularization used by the model to minimize
overfitting, and $||\cdot||_F$ is the Frobenius norm.

\section{\taskOne}\label{sec:feature}
Traditional methods usually leverage historical user-POI interactions
by mapping users and POIs to a shared latent space via factorization
over a user-POI affinity matrix.
However, the learned latent space rarely provides any insight into
why a user prefers a POI~\cite{zhang2014explicit}.
Worse still, such data is often quite
sparse~\cite{hang2018exploring}, which may not be sufficient to
provide meaningful signals.

To address this limitation, we leverage the attributes of both
users and POIs, which provide complimentary evidences (i.e., the 
``\explainFeature hint'' introduced in Section~\ref{sec:explain}) to
reveal to a user why certain POIs are being recommended.
This allows a user to interactively provide additional information to
align the current recommendations with their information need.
We refer to these attributes that can be directly derived from the
dataset as \textit{explicit features}, e.g., user's age and hobby.
In contrast, \textit{implicit features} correspond to the attributes
inferred from available data.
To this end, we learn an attribute-based representation for our
recommender system.

\vspace{-2ex}
\subsection{Objective Loss Function} \label{subsec-loss1}
Before introducing details on model training using the above
attributes, we define the first loss function to be used in our
approach.
Similar to previous matrix factorization models for user-POI check-in
records, we derive a factorization model over the observed
user-attribute matrix $\userFeatureMatrix \in \mathbb{R}^{m \times
\featureNum}$ and POI-attribute matrix $\poiFeatureMatrix \in
\mathbb{R}^{\poiNum
\times \featureNum}$ to learn \textit{explicit feature}
representations for users and POIs, where \featureNum is the total number
of explicit features of users and POIs.
This can be achieved by minimizing the following loss function:
\vspace{-1.5ex}
\begin{align}
	&\lossAttribute = ||\userWithHiddenMatrix \featurelatent-\userFeatureMatrix||_F^2+ \sum_{l=1}^{\poiTreeNum}||\poiWithHiddenMatrix \featurelatent-\poiFeatureMatrix||_F^2
	\label{equ-loss1}\vspace{-1ex} 
\end{align} 
where $\userWithHiddenMatrix \in \mathbb{R}^{m \times r}$ and
$\poiWithHiddenMatrix \in \mathbb{R}^{\poiNum \times r}$ are two
learned parameters to model the explicit feature representations of
users and POIs, which are then combined with a shared latent vector
$\featurelatent \in \mathbb{R}^{r \times \featureNum}$.
Here, $r$ is the latent factor magnitude.

\vspace{-1ex}
\subsection{Representation of User and POI Features}
\label{subsec-userfeature}
We first show how to build the matrix {\userFeatureMatrix} to
incorporate the attribute values of users.
\userFeatureMatrix is in turn a concatenation of two matrices
capturing different contexts -- (1) a \textit{direct attribute matrix
\userAttMatrix} directly obtained from user attributes; (2) an
\textit{inverse attribute matrix \userTagMatrix} induced from
the empirical user check-in distribution:
\begin{equation}
\label{equ-userfeature}
\userFeatureMatrix = \userAttMatrix \oplus \userTagMatrix
\end{equation}
where $\userAttMatrix \in \mathbb{R}^{m \times \featureNumUser}$,
$\userTagMatrix \in \mathbb{R}^{m \times \featureNumPoi}$,
$\userFeatureMatrix \in \mathbb{R}^{m \times \featureNum}$,
$\featureNum=\featureNumUser+\featureNumPoi$, and $\oplus$ is the
concatenation operator.

Similarly, we construct the attribute matrix \poiFeatureMatrix for
POIs at the $l$-th level of \poiTree, which in turn is a
concatenation of a \textit{direct attribute matrix} \poiAttMatrix and an \textit{inverse attribute matrix} \poiTagMatrix:
\begin{equation}
\label{equ-poiFeature}
\poiFeatureMatrix = \poiAttMatrix \oplus \poiTagMatrix
\end{equation}
where $\poiAttMatrix \in \mathbb{R}^{n_l \times \featureNumPoi}$,
$\poiTagMatrix \in \mathbb{R}^{n_l \times \featureNumUser}$,
$\poiFeatureMatrix \in \mathbb{R}^{n_l \times \featureNum}$.
%
We use \featureNumUser and \featureNumPoi to denote the number of
user features and POI features generated from their respective attributes. The concatenation process is illustrated in the lower left corner of Figure \ref{fig-framework}. 

\myparagraph{Constructing the direct attribute matrix}
Raw attribute values can be numerical (e.g., the age is $18$) or
binary (e.g., a hobby such as \textit{reading}).
We empirically define various decision rules to split an attribute $a_k$
into two decision features.
For any numerical attribute (e.g., age), a threshold
$\theta_k$ is selected to split the attribute into $[a_k \textless
\theta_k]$ and $[a_k \geq \theta_k]$.
Note that, multiple threshold values can also be used to
split one attribute empirically, which generates a corresponding number of features.
For a binary attribute (e.g., country), we have $[a_k = \theta_k]$ or
$[a_k \neq \theta_k]$.
\begin{equation}\label{equ-xa}
	\text{\userAttMatrix}_{i,k} = \left\{\begin{matrix}
	1 & \text{If \userOne satisfies the decision rule over }a_k	\\ 
	0 & \text{Otherwise}
	\end{matrix}\right.
\end{equation}

Given the attribute set of users and the attribute set of POIs
located at the $l$-th level of \poiTree, we model the
\textit{direct attribute matrices} \userAttMatrix (Eq.
\ref{equ-xa}) and \poiAttMatrix (Eq.
\ref{equ-ya}) as a concatenation of one-hot vectors, where
an element of value $1$ denotes a fulfilled decision rule.
\begin{equation}\label{equ-ya}
\text{\poiAttMatrix}_{j,k} = \left\{\begin{matrix}
1 & \text{If \poiOne satisfies the decision rule over }a_k	\\ 
0 & \text{Otherwise}
\end{matrix}\right.
\end{equation}

\myparagraph{Constructing the inverse attribute matrix}
We assume that users visit only the venues they are interested in, e.g.,
if Alice often goes to the library, she may be a book-lover.
However, such info has to be inferred as it may not be available in
the user profile (hobbies).
This assumption allows us to enrich the raw data, and is a form of
{\em weak supervision} \cite{dehghani2017neural}.
Leveraging the attributes of POIs visited by users in this manner
somewhat mitigates sparsity and cold-start issues commonly
encountered in recommendation modeling.

If a POI \poiOne, which has an attribute $a_k$ and was visited by a
user \userOne for \userTagTimes times, then $tp_{ik} = \sum tp_{ijk}$ and each element in the user \textit{inverse attribute matrix} \userTagMatrix is computed as follows (assume min-max normalization):
\begin{equation}\label{equ-xt}
	\text{\userTagMatrix}_{i,k} = \left\{\begin{matrix}
	\frac{tp_{ik} - \userTagTimesMin}{\userTagTimesMax - \userTagTimesMin} & \text{If \userOne visited \poiOne that has }a_k	\\ 
	0 & \text{Otherwise}
	\end{matrix}\right.
	\vspace{-1ex}
\end{equation}
where \userTagTimesMax and \userTagTimesMin are the highest and
lowest check-in frequency for \userOne, respectively.

Similarly, attributes for the users who checked in a specific POI
\poiOne represent the \textit{inverse attributes}.
Suppose a POI was visited by \poiAttTimes users who have an attribute
$a_k$, then $tu_{jk} = \sum tu_{ijk}$ and each element in the POI \textit{inverse attribute matrix}
\poiTagMatrix is:
\begin{equation}\label{equ-yt}
	\text{\poiTagMatrix}_{j,k} = \left\{\begin{matrix}
	\frac{tu_{jk} - \poiAttTimesMin}{\poiAttTimesMax - \poiAttTimesMin} & \text{If \poiOne was visited by \userOne who has }a_k 	\\ 
	0 & \text{Otherwise}
	\end{matrix}\right.
\end{equation}
where \poiAttTimesMax and \poiAttTimesMin are the largest and the
smallest number of users who visit \poiOne, respectively.

\section{\taskTwo}\label{sec:interaction}
In this section, we will show how to further boost the recommendation
performance by exploiting user-POI interactions.

\vspace{-1ex}
\subsection{Objective Loss Function}\label{subsec-loss2}
We leverage the Bayesian Personalized Ranking
(BPR)~\cite{rendle2009bpr} principle to construct the loss function
\lossInteraction for the second subtask.
Specifically, following the popular negative sampling strategy
\cite{li2015rank,wang2018streaming}, a negative POI instance \poiNeg
which the user never visited is paired with a positive POI instance
\poiPos, and the pairwise log loss can be computed by maximizing the
difference between the prediction scores of the positive and negative
samples.
\lossInteraction is shown as follows:
\begin{equation}
	\vspace{-0.2em}
	\label{equ-loss2}
	\lossInteraction = - \sum_{l=1}^{\poiTreeNum} \sum_{i=1}^{m} \sum_{j=1}^{\poiNum} \ln \sigma (\scorefinal_{i, \poiPos} - \scorefinal_{i, \poiNeg})
	\vspace{-0.2em}
\end{equation}
where $\scorefinal_{i, \poiPos}$ (or $\scorefinal_{i, \poiNeg}$) is
the predicted score w.r.t.\ a positive POI \poiPos (or a negative POI
\poiNeg) located at the $l$-th POI level for the $i$-th user.
Here, we add the \textit{minus} sign in the front to match the
minimization objective with Eq.
\ref{equ-lossShort}.
The \emph{user-POI check-in matrix} $\scorefinal \in \mathbb{R}^{m
\times n_l}$ will be further elaborated next.

\vspace{-1ex}
\subsection{Modeling user-POI interaction}\label{subsec-interMatrix}
We incorporate two matrices $\predScoreInter \in \mathbb{R}^{m \times
\poiNum}$ and $\predScoreHis \in \mathbb{R}^{m \times \poiNum}$ into
\scorefinal through a linear combination, where
\predScoreInter denotes the \textit{feature-based check-in matrix}, and
\predScoreHis is the \textit{historical check-in matrix}.
A configurable parameter \adjustPara is used to control
the relative contributions of these two matrices, resulting in the
following equation:
\vspace{-0.2em}
\begin{equation}
\label{equ-pred}
\scorefinal = \predScoreInter + \adjustPara \predScoreHis
\vspace{-0.3em}
\end{equation}
By combining \predScoreInter and \predScoreHis, we obtain the final
top-$k$ recommended results sorted by similarity score \scorefinal.
This process is illustrated in the lower right corner of
Figure \ref{fig-framework}.
Next, we show how to construct \predScoreInter and {\predScoreHis}.
\vspace{-0.5em}
\subsubsection{Constructing the feature-based check-in matrix}\label{subsubsec-feature_checkin}
In order to fully leverage the interaction data of users and POIs,
the feature-based check-in matrix \predScoreInter located at the
$l$-th level is built based on the feature representation \matrixP
and \matrixQ w.r.t.\ users and POIs, respectively:
\begin{equation}
\label{equ-predScoreInter}
\vspace{-0.3em}
\begin{split}
&\predScoreInter=\matrixP (\matrixQ)^T, \matrixP = \userWithHiddenMatrix \oplus \userHiddenMatrix \oplus
\userLevelMatrix, \matrixQ = \poiWithHiddenMatrix \oplus \poiHiddenMatrix \oplus \poiLevelMatrix	\\
\end{split}
\vspace{-0.3em}
\end{equation}

In Eq.\ \ref{equ-predScoreInter}, $\matrixP \in \mathbb{R}^{m \times
(r+r_l+r_{l+1})}$ is a concatenation of three matrices w.r.t.
users: \userWithHiddenMatrix, \userHiddenMatrix, and
\userLevelMatrix.
Specifically, \userWithHiddenMatrix is the explicit feature
representation of users.
$\userHiddenMatrix \in \mathbb{R}^{m \times r_l}$ is the implicit
feature representation of users, and $\userLevelMatrix \in
\mathbb{R}^{m \times r_{l+1}}$ is a trainable matrix parameter to
match \poiLevelMatrix in the same space.
Here, $r_l$ denotes the latent factor size of implicit features at
the $l$-th level of \poiTree.

Accordingly, $\matrixQ \in \mathbb{R}^{n_l \times (r+r_l+r_{l+1})}$
incorporates three kinds of information w.r.t.\ the POIs at the
$l$-th level of \poiTree: \poiWithHiddenMatrix is the explicit
feature representation, $\poiHiddenMatrix \in \mathbb{R}^{n_l
\times r_l}$ is the implicit feature representation, and
$\poiLevelMatrix \in \mathbb{R}^{n_l \times r_{l+1}}$ is the
inter-level POI feature representation propagated from child POIs
with an attention network.

Recall that \userWithHiddenMatrix and \poiWithHiddenMatrix were
described in Section~\ref{sec:feature}.
We now describe the details on how to construct implicit feature
representations \userHiddenMatrix and \poiHiddenMatrix, and how to
produce an inter-level POI feature representation \poiLevelMatrix.

\myparagraph{Implicit feature representation}
Some features that influence user preferences may be implicit.
For example, Alice might go to historical libraries because she loves
the classical architecture there, or for other unknown reasons which
cannot be inferred.
These types of features can be learned by using two matrices
{\userHiddenMatrix} and {\poiHiddenMatrix} w.r.t.\ users and POIs,
respectively.

\myparagraph{Inter-level propagated POI feature representation}
The feature information covered by a child POI can also be used by
its parent POI.
For instance, the attributes of child POIs (e.g., a restaurant or a
store) can be aggregated into its parent POI (e.g., a mall).

In particular, for each parent POI \poiAttParent, we also propagate a
learned implicit feature representation (i.e., an embedding vector
\poiImplictEmbeddingChild in $\vec{H}_p^{l+1}$) from each child POI
\poiAttChild to \poiAttParent, producing the inter-level feature
representation \poiAttImplicitEmbeddingParent for
\poiAttParent to leverage the inter-level information.
Here, we denote $\poiLevelMatrix \in \mathbb{R}^{n_l \times r_{l+1}}$
as the inter-level POI feature representation matrix for all POIs at
the $l$-th level of \poiTree, where \poiAttImplicitEmbeddingParent is
an embedding vector for a POI in \poiLevelMatrix.
Next, we show how to induce \poiAttImplicitEmbeddingParent in detail.

One possible way to learn \poiAttImplicitEmbeddingParent is to
augment the implicit features in all its child POIs.
However, different child POIs might provide different contributions
when influencing the parent POIs.
For example, many users may visit a shopping mall (a parent POI)
frequently for a popular grocery store (a child POI) and nothing
else.

To mitigate this issue, we propagate learned implicit features from a
child POI \poiImplictEmbeddingChild using various attention weights
throughout \poiTree in order to learn the best inter-level feature
representation \poiAttImplicitEmbeddingParent for a parent POI
\poiAttParent.
Specifically, we use a multi-layer perceptron (MLP) when learning
attention weights each child POI \poiTreeNodeChildren rooted at
\poiTreeNodeParent.
\begin{equation}
	\vspace{-0.5em}
	\left\{\begin{matrix}
	w_j^{l+1}=F(\poiImplictEmbeddingChild)= ReLU(\vec{d} (\poiAttentionWeight\poiImplictEmbeddingChild+\vec{b_1}))+b_2	\\
	\tilde{w}_j^{l+1}=\sigma(w_j^{l+1})=\frac{exp(w_j^{l+1})}{\sum_{p_t^{l+1}\in C(\poiTreeNodeParent)}exp(w_t^{l+1})} 	\\ 
	\poiAttImplicitEmbeddingParent=\sum_{\poiTreeNodeChildren \in C(\poiTreeNodeParent)}\tilde{w}_j^{l+1}\poiImplictEmbeddingChild
	\end{matrix}\right.
	\vspace{-0.25em}
\end{equation}
where the \textit{implicit feature} embedding \poiImplictEmbeddingChild $\in
\mathbb{R}^{r_{l+1}}$ of child POI \poiAttChild is 
the input, and $ReLU(x)=max(0,x)$ is applied as the activation
function to produce $w_j^{l+1}$ in the first formula.
$\poiAttentionWeight \in \mathbb{R}^{d_1 \times r_{l+1}}$ is a
transpose matrix, $\vec{b_1} \in \mathbb{R}^{d_1}$ denotes a bias
vector, $b_2$ refers to a bias variable, and $\vec{d} \in
\mathbb{R}^{1 \times d_1}$ projects the attention weight for a POI node where
the hidden layer size of the attention network is
\poiAttentionHidden. 
$C(\poiTreeNodeParent)$ indicates all child POIs rooted at \poiTreeNodeParent. 

After computing the attention weight $w_j^{l+1}$, we normalize it to
obtain $\tilde{w}_j^{l+1}$ using a softmax function $\sigma(\cdot)$
as shown in the second formula.
Finally, \poiAttImplicitEmbeddingParent is produced using the
resulting child POIs and attention weights in the third formula.
The complete architecture of our attention network mechanism is
depicted in the centre of Figure~\ref{fig-framework}.

\vspace{-1ex}
\subsubsection{Constructing historical check-in matrix}\label{subsubsec-history_checkin}
Intuitively, a POI candidate may be recommended if it is located near
a previously visited POI.
To exploit spatial containment, we first construct
\poiTreeNum POI context graphs, one for each POI level.
Each POI context graph embeds the contextual information of the POIs.
The mechanism used to incorporate contextual information between a
POI candidate and a visited POI into our recommendation model is
described next.

\myparagraph{POI context graph}
For ease of illustration, we use a single POI context graph as an
example and omit superscripts (i.e., $l$) when denoting a particular
level in \poiTree.
Specifically, we represent a POI context graph as
$\poiGeospatialGraph=\langle \poiGeospatialVertex, \poiGeospatialEdge
\rangle$, where \poiGeospatialVertex is the set of POIs, and
\poiGeospatialEdge is the set of edges between any two connected
POIs.
Given any two POIs \poiExampleOne and \poiExampleTwo
($\poiExampleOne, \poiExampleTwo \in \poiGeospatialVertex$), we
define three types of edge relations, such that \poiGeospatialEdge
can be further weighted using multiple geospatial influence factors.
\begin{itemize}[leftmargin=*]
\item \textit{Co-search.} If a user searches for a restaurant and a 
coffee shop within a short time interval using a map application, and
then visits the restaurant, we can infer that a coffee shop has
a higher likelihood of relevance the next time the user views the map
\cite{zhou2018early}.
Thus, we use \poiExampleCoQuery to denote the co-occurrence search
frequency between two POIs \poiExampleOne and \poiExampleTwo within a
fixed session interval $\Delta t_1$ (e.g., 30 minutes) for all users.
\item \textit{Co-visit.} If a user first visits a restaurant and then 
goes to a coffee shop and locations are being tracked for the user,
we assume that the coffee shop has a higher priority for
recommendations made when a user is located in a restaurant.
We use \poiExampleCoVisit to represent the visit frequency
chronologically between \poiExampleOne and \poiExampleTwo within a
fixed time interval $\Delta t_2$ (e.g., 30 minutes).
\item \textit{Geospatial distance.} 
According to Tobler’s first law of geography
\cite{tobler1970computer}, ``everything is related to everything
else, but near things are more related than distant things''.
The nearby objects often have underlying relationships and influence,
thus we also apply a \textit{geospatial distance} factor which
captures the geographical influence.
Here, we use \poiExampleCoDist to denote the inverse Euclidean
distance between \poiExampleOne and \poiExampleTwo.
\end{itemize} 

Note that \poiGeospatialGraph is constructed before training.
The edge weights derived using these three geospatial factors are
normalized using sigmoid function, which is defined as
$\sigma(x)=1/(1+exp(-x))$.

\myparagraph{Graph-based geospatial influence
representation}\label{subsubsec-georep} Given a POI candidate $p_i$
to be recommended and a historical POI check-in trajectory
$\checkinSeq^{u_k}$ for a user $u_k$, we define the geospatial influence matrix $U_{g, u_k}^l \in
\mathbb{R}^{\poiNum \times r}$, and incorporate POI context
info using Eq.~\ref{equ-poi-spatial}.
Since using every visited POI from $\checkinSeq^{u_k}$ is not scalable, we
only choose a subset $\checkinSeqSubset^{u_k}$ containing the top-$t$
frequently visited POIs from $\checkinSeq^{u_k}$ for each user $u_k$ such that $\checkinSeqSubset^{u_k}
\subset \checkinSeq^{u_k}$ and $|\checkinSeqSubset^{u_k}| = t$.
Specifically, we denote the embedding vector for $p_i$ in $U_{g, u_k}^l$ as $\mu_{g, u_k}^{l, p_i}$, and the embedding vector for $p_j$ in $U_p^l$ as $\upsilon_{p_j}^l$. Thus for the recommended POI $p_i$ and the historical visited POI $p_j$, we can get:
\begin{equation}
	\label{equ-poi-spatial}
	\mu_{g, u_k}^{l, p_i} = \frac{1}{t} \sum_{p_j \in \checkinSeqSubset^{u_k}}^{} \poiCoQuery\poiCoVisit\poiCoDist \upsilon_{p_j}^l
\end{equation}
where $t$ is set to $3$ in our experiment.
Consequently, the embedding vector $g_{u_k}^l$ for the user $u_k$ in the historical check-in matrix \predScoreHis can be computed as:
\vspace{-.5em}
\begin{equation}
\label{equ-pred-spatial}
g_{u_k}^l = \omega_k  (U_{g, u_k}^l)^T
\end{equation}
where $\omega_k$ is an embedding vector for $u_k$ in $U_u$. Finally, \predScoreHis can be built as $\predScoreHis = [\dots, (g_{u_k}^l)^T. \dots]^T$ for all users.   

Note that POI recommendation task can be formalized as a
top-$k$ ranking problem. Once we have learned the model parameters in
\modelnameshort, given a user, a ranking score for each POI located
at the $l$-th level of \poiTree can be obtained from the
matrix \scorefinal, and then the  POIs with top-$k$ highest ranking
scores will be recommended to the user.

\vspace{-2ex}
\section{Hints for Recommendation Justification} \label{sec:explain}
It is desirable to complement recommendations with an intuition as to
why certain results are produced, since it may not always be
obvious to the user~\cite{zhang2018explainable}.
Our approach provides such additional benefit by enabling (i)
\textit{\explainFeature} hint: user attributes used by the model can
be derived; (ii) \textit{\explainPOI} hint: when a parent POI is
recommended, specific child POIs can be discovered; and (iii)
\textit{\explainHistory} hint: if we recommend a new POI, we can
highlight data from historical check-in venues that were most
relevant.

\myparagraph{User-aspect} 
We assume that a user \userOne has visited a POI
\poiOne based on the attributes of that POI.
Our model captures the top-$K$ features for \userOne from an explicit
feature embedding vector $\vec{uf}$, obtained from a row vector from
$\vec{M}_u$ matrix, which is computed by
$\vec{M}_u=\userWithHiddenMatrix \featurelatent$ (as mentioned in
Section \ref{sec:feature}).
$K$ is set to $5$ in our experiment.
Thus, the column index set $B_i = \langle b_{i1}, b_{i1}, ..., b_{iK}
\rangle$ are the top-$K$ ranked in $\vec{uf}$.
The matrix $\vec{M}_p^l=\poiWithHiddenMatrix
\featurelatent$ is used to determine the POI explicit feature
embedding vector $\vec{pf}^l$ and find the corresponding POI feature
prediction values based on $B_i$.
We can then expose the POI feature with the highest value to \userOne
for recommendation evidence.
\vspace{-1em}
\begin{figure}[h]
	\centering
	\includegraphics[width=6cm]{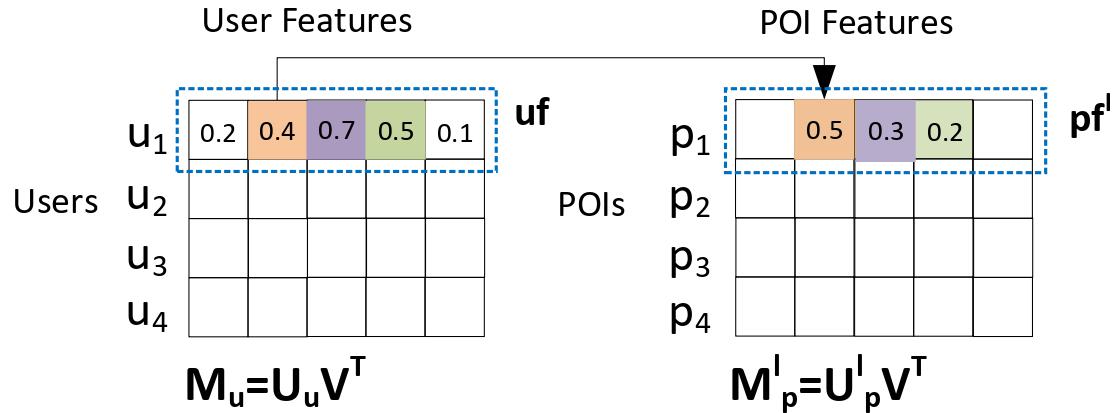}
	\vspace{-2.5ex}
	\mycaption{Illustration example on user-aspect hint.}
	\label{fig-hint_first}
	\vspace{-2.5ex}
\end{figure}
\vspace{-.3em}

An illustrative example of a user-aspect hint is shown in Figure
\ref{fig-hint_first}.
After obtaining the two matrices $\vec{M}_u$ and $\vec{M}_p^l$, say
for the user $u_1$, the user feature with the highest $K$ values
(assuming that $K=3$, then $B_1 =\langle 2,3,4 \rangle$) is located
in the second, third, and fourth column using the embedding
$\vec{uf}$.
Then, the corresponding POI features whose column indexes drop into
$B_1$ are identified, where the POI feature with the highest value
$0.5$ in the second column of $\vec{pf}^l$ can then be presented as a
hint to the user.

\myparagraph{POI-aspect}
Intuitions about parent POI recommendations can be derived from the
attention influence weights computed for each child POI (as described
in Section~\ref{subsubsec-feature_checkin}).
If we recommend a parent POI $p$ to a user \userOne, a set of
important child POIs can be shown, ordered by attention scores.
Thus the contribution ratio for each child POI \poiOne $(\poiOne \in
C(p))$ over all child POIs $C(p)$ is computed by
$\frac{\vec{a_{\userOne} \odot \vec{a_{\poiOne}}}}{ \sum_{p_c \in
C(\poiOne)} \vec{a_{\userOne}} \odot \vec{a_{p_c}} }$, where
$\vec{a_{\userOne}}$ is a user embedding in \userLevelMatrix,
$\vec{a_{\poiOne}}$ and $\vec{a_{p_c}}$ are two POI embedding vectors
in \poiLevelMatrix, and $\odot$ is the dot product
operator.
We mark the child POI with the highest contribution ratio as a
``hot'' POI which might attract the user.

\myparagraph{Interaction-aspect}
For any recommended POI $p_j$, we can easily evaluate the contribution to examine whether the historical check-in information influences the final prediction.
We define the contribution ratio as the prediction score
$\predScoreHis_{\userOne, p_j}$ (as introduced in
Section~\ref{subsubsec-history_checkin}) on historical interactions
divided by the total predicted score $\vec{O}^l_{\userOne, p_j}$,
which is $\expParaHis =
\frac{\predScoreHis_{\userOne, p_j}}{\vec{O}^l_{\userOne,
p_j}}$.
If $\expParaHis$ exceeds a threshold, we assume that the historical check-in information is the important contributor when recommending $p_j$ to \userOne. 

\section{Experimental study}\label{sec:exp}
We investigate the following four research questions:
\begin{itemize}[leftmargin=1em]
\item \noindent \textbf{RQ1.} How does our proposed \modelnameshort 
model perform when compared with the state-of-the-art POI
recommendation methods?
\item \noindent \textbf{RQ2.} How does \modelnameshort perform when 
varying the hyper-parameter settings (e.g., \textit{embedding size})?
\item \noindent \textbf{RQ3.} How can \modelnameshort be used to provide 
recommendation hints?
\item \noindent \textbf{RQ4.} How do different components in \modelnameshort 
contribute to the overall performance?
\end{itemize}

We evaluate all methods using two real-world city-wide datasets,
\textit{Beijing} and \textit{Chengdu}, from Baidu
Maps\footnote{https://map.baidu.com}, which is one of the most
popularly used map services in China.
Both datasets are randomly sampled as a portion of whole data from
Baidu Maps.
Due to space limitations, we only show the experimental results for
the \textit{Beijing} dataset, except when answering \textbf{RQ1}.
Similar performance trends were observed for the \textit{Chengdu}
dataset when answering RQ2-RQ4.
\begin{itemize}[leftmargin=*]
\item \textit{The POI tree \poiTree.}
We trace the profile for each POI and then recursively search its
parent POI to build \poiTree.
A three-level POI tree is built: \poiTreeLayerOne, \poiTreeLayerTwo,
and
\poiTreeLayerThree from top to bottom.
For example, a spatial containment path in \poiTree on the
\textit{Beijing} dataset is \textit{Wudaokou} (a famous neighborhood
in \textit{Beijing})$\rightarrow$\textit{Tsinghua
University}$\rightarrow$\textit{Tsinghua Garden}, which are located
at \poiTreeLayerOne, \poiTreeLayerTwo, and \poiTreeLayerThree,
respectively.

\item \textit{Check-in data.}
Each check-in has the following info: userId, poiId and a check-in
timestamp.
We filter out users with fewer than $10$ check-in POIs and POIs
visited by fewer than $10$ users.
To build the check-in data on \poiTreeLayerOne and
\poiTreeLayerTwo, the check-in records from users was used and we
also aggregated the check-ins in the parent POIs if any of their
child POIs were visited.

\item \textit{User and POI profile.}
Each user has their own attributes such as age and hobby, and
\featureNumUser=$173$ user features are extracted.
Each POI has a parent POI, and its own attributes, where
\featureNumPoi=$467$ representative POI features are available after
filtering out those attributes shared by fewer than $10$ POIs.
\end{itemize}

\myparagraph{Setup}
We partitioned the check-in data into a training set, a validation
set, and a test set.
The first two months of check-ins were used for training in the
\textit{Beijing} testset, and the first three months in \textit{Chengdu}.
The most recent $15$ days of check-ins were used as the test data and
all remaining ones were used in the validation data in both datasets.
A negative sample was randomly selected for each positive sample
during training.
Any check-in that occurred in the training set was pruned
from both the validation and test set, to ensure that any POI
recommended had never been visited by the user before.

For each model, the parameters were tuned on the validation data to
find the best values that maximized $P@k$, and used for all test
predictions.
Mini-batch adaptive gradient descent~\cite{duchi2011adaptive} is used
to control the learning step size dynamically.
All experiments were implemented in Python on a GPU-CPU platform
using a GTX 1080 GPU.

\myparagraph{Evaluation Metrics}
We adopt two commonly-used performance metrics
\cite{liu2017experimental}: Precision ($P@k$), and
Normalized Discounted Cumulative Gain ($NDCG@k$).
These two metrics were used to evaluate the model performance since
$P@k$ is commonly used when evaluating the coverage of recommendation
results, and $NDCG@k$ captures additional signals about the overall
effectiveness of the top-$k$ recommendations, and supports graded
relevance.

\myparagraph{Parameter Settings}
The parameters $\Delta t_1$ and $\Delta t_2$ are set to $30$ minutes
by default.
The adjustable parameter \adjustPara for graph-based geospatial
influence is set to $1$ by default, and the regularization parameters
are set as follows: $\lambda_1=0.01$ and $\lambda_2=0.1$, both of
which are set according to the experiment evaluation using the
validation dataset.
Furthermore, the hidden factor size $r_{l}$ of the POI levels are
fixed, and we empirically set the attention layer size $d_1$ to be
the same as $r_{l}$, which is equal to $150$ discovered during the
parameter tuning experiment shown in Table~\ref{table-para-control}.
\vspace{-0.5em}
\subsection{Overview}


\newcolumntype{P}[1]{>{\centering\arraybackslash}p{#1}}
\begin{table*}[t]
	\footnotesize	
	\renewcommand{\arraystretch}{1.2}
	\mycaption{Model performance comparisons on the \textit{Beijing} and \textit{Chengdu} dataset. Entries marked \tup and \btup correspond to statistical significance using a paired t-test with Bonferroni correction at $95\%$ and $99.9\%$ confidence intervals respectively. Comparisons are relative to PACE.}
	\vspace{-3ex}
	\label{table-k-beijing}
	\centering
	\begin{tabular}{cccccccccccccc}
		\toprule
		\multirow{2}{*}{Level} & \multirow{2}{*}{Model} & \multicolumn{6}{c}{\textit{Beijing}} & \multicolumn{6}{c}{\textit{Chengdu}} \\
		\cmidrule{3-14}
		&                   & P@5         & NDCG@5      & P@10      & NDCG@10	& P@20       & NDCG@20 & P@5        & NDCG@5      & P@10        & NDCG@10 & P@20      & NDCG@20  \\ \midrule 
		\multirow{5}{*}{$H_1$} 
		& WRMF 		&0.056\btdown	&0.096\btdown	&0.047\btdown	&0.121\btdown	&0.037\btdown	&0.151\btdown	&  0.063\btdown&	0.079\btdown&	0.051\btdown&	0.098\btdown&	0.041\btdown&	0.127\btdown \\  
		& BPRMF		&0.079\btup	&0.123\btup	&0.064\btup	&0.150\btup&	0.050\btup&	0.187\btup &0.110\btup&	0.142\btup&	0.086\btup&	0.170\btup&	0.061\btup&	0.202\btup  \\ 
		& PACE 		&0.067\;\;  &0.104\;\;	&0.053\;\;	&0.124\;\;&	0.043\;\;&	0.156\;\;  & 0.087\;\;&	0.117\;\;&	0.074\;\;&	0.152\;\;&	0.054\;\;&	0.181\;\; \\ 
		& SAE-NAD	&0.078\btup	&0.125\btup	&0.064\btup	&0.155\btup&	0.051\btup&	0.194\btup   & 0.100\btup&	0.128\btup&	0.081\btup&	0.155\btup&	0.057\btup&	0.185\btup  \\ 
		& MPR		&\textbf{0.084}\btup&	\textbf{0.133}\btup&	\textbf{0.067}\btup&	\textbf{0.162}\btup&	\textbf{0.053}\btup&	\textbf{0.203}\btup  & \textbf{0.119}\tup&	\textbf{0.159}\tup&	\textbf{0.094}\btup&	\textbf{0.190}\btup&	\textbf{0.064}\btup&	\textbf{0.222}\btup
		\\ \midrule 
		
		\multirow{5}{*}{$H_2$} 
		& WRMF &0.009\;\; &	0.017\;\;&	0.007\;\;&	0.022\;\;&	0.005\;\;	&0.026\btup & 0.022\;\;&0.027\;\;	&0.018\tdown&	0.034\;\;&	0.013\;\;&	0.040\;\;  \\ 
		& BPRMF &0.007\;\;&	0.014\btup&	0.007\;\;&	0.020\btup&	0.005\;\;&	0.026\btup & 0.027\;\;&	0.037\;\;&	0.022\;\;&	0.047\;\;&	0.017\;\;&	0.058\;\; \\ 
		& PACE  &0.007\;\;&	0.013\;\;&	0.007\;\;&	0.019\;\;&	0.005\;\;&	0.024\;\; &0.022\;\;&	0.031\;\;&	0.022\;\;&	0.039\;\;&	0.013\;\;&	0.046\;\; \\ 
		& SAE-NAD & 0.007\;\;&	0.014\btup&	0.006\btdown&	0.018\btdown&	0.005\;\;&	0.024\;\; & \textbf{0.033}\;\;&	0.043\;\;&	0.019\;\;&	0.049\;\;&	0.017\;\;&	0.059\;\;  \\ 
		& MPR &\textbf{0.010}\btup&	\textbf{0.018}\btup&	\textbf{0.008}\btup&	\textbf{0.023}\btup&	\textbf{0.007}\btup&	\textbf{0.030}\btup & \textbf{0.033}\btup&	\textbf{0.044}\btup&	\textbf{0.026}\tup	&\textbf{0.054}\btup&	\textbf{0.020}\btup	&\textbf{0.067}\btup
		\\ \midrule 
		
		\multirow{5}{*}{$H_3$} 
		& WRMF & 0.008\btup&	0.015\btup&	0.006\btup	&	0.018\btup&	0.004\;\;&	0.022\btup &  0.021\tup&0.027\tup&	0.017\;\;	&0.033\;\;	&0.013\;\;	&0.041\;\;	 \\ 
		& BPRMF & 0.006\btdown&	0.012\btup&	0.005\;\;&	0.015\btup&	0.004\;\;&	0.019\btup & 0.021\tup&	0.029\;\;&	0.017\;\;&	0.036\;\;&	0.013\tup&	0.043\btup  \\ 
		& PACE & 0.007\;\;	&	0.008\;\;&	0.005\;\;&	0.009\;\;&	0.004\;\;&	0.010\;\; &  0.016\;\;&	0.023\;\;&	0.016\;\;&	0.032\;\;&	0.009\;\;&	0.035\;\;\\ 
		& SAE-NAD & 0.008\btup&	0.015\btup&	0.007\btup	&	0.020\btup&	0.005\btup&	\textbf{0.026}\btup & 0.020\btup&	0.027\btup&	0.020\tup&	0.038\btup&	0.016\btup&	0.047\btup \\ 
		& MPR &\textbf{0.009}\btup&	\textbf{0.015}\btup&	\textbf{0.007}\btup &	\textbf{0.021}\btup&	\textbf{0.006}\btup&	\textbf{0.026}\btup & \textbf{0.032}\btup&	\textbf{0.042}\btup&	\textbf{0.021}\btup&	\textbf{0.046}\btup&	\textbf{0.016}\btup&	\textbf{0.056}\btup
		\\ \bottomrule 
	\end{tabular}
	\vspace{-2ex}
\end{table*}

\myparagraph{Baselines}
To validate the performance of our model \modelnameshort, we compared
directly against the following state-of-the-art methods.
Note that, these baselines all treat POIs as isomorphic, thus we have
to construct multiple models, one for each POI level, in order to
generate comparable output to our approach.
\begin{itemize}[leftmargin=*]
\item WRMF (Weighted Regularized Matrix
Factorization)~\cite{hu2008collaborative}: a point-wise latent factor
model that distinguishes user observed and unobserved check-in data by
using confidence values to adapt to implicit feedback data from a user.
\item BPRMF (Bayesian Personalized Ranking)~\cite{rendle2009bpr}: a
pair-wise learning framework for implicit feedback data,
combined with matrix factorization as the internal predictor.
\item PACE (Preference and Context
Embedding)~\cite{yang2017bridging}: a neural embedding approach that 
generally combines user check-in behaviors and context information
from users and POIs through a graph-based semi-supervised learning
framework.
\item SAE-NAD (Self-attentive Autoencoders with Neighbor-Aware
Influence)~\cite{ma2018point}: explicitly integrates spatial
information into an autoencoder framework and uses a self-attention
mechanism to generate user representation from historical check-in
records.
\end{itemize}

\vspace{-2ex}
\subsection{Effectiveness Comparisons (RQ1)}
\subsubsection{Baseline Comparisons}
Table~\ref{table-k-beijing} compares all methods using different $k$
values on both datasets.
The key observations can be summarized as follows.
\begin{itemize}[leftmargin=1em]
\item Our model \modelnameshort achieves the best performance on all
metrics at every single level of spatial granularity, demonstrating
the robustness of our model.
Specifically, the NDCG@10 for \modelnameshort on \textit{Beijing}
has: (1) a $4.5\%$ improvement over the best baseline SAE-NAD at the
\poiTreeLayerOne level; (2) a $4.5\%$ improvement over the strongest
baseline WRMF at the \poiTreeLayerTwo level; and
(3) a $5\%$ improvement over the best baseline SAE-NAD at the 
\poiTreeLayerThree level.
\item In term of P@10, \modelnameshort substantially outperforms WRMF
and BPRMF ($42.6\%$ and $4.7\%$ respectively) at the \poiTreeLayerOne
level.
This results from WRMF and BPRMF treating each POI level
independently when training the model.
Clearly, \modelnameshort benefits from jointly optimizing the loss
for every level of \poiTree in order to achieve its collaborative
training goal.
\item Both PACE and SAE-NAD directly incorporate geospatial distance
information.
We can observe that SAE-NAD outperforms PACE in most cases.
One potential reason is that although PACE builds a context graph to
model important geographical influences, it ignores the
historical visit information when extracting the POI-POI co-visit
relations that are used by SAE-NAD.
However, SAE-NAD employs an autoencoder and self-attention mechanism
when constructing POI-and-POI relations, while our model \modelnameshort
is able to learn the geospatial influence relations across all
\poiTree levels, and the additive benefits are clear.
As such, we believe that the inter-level relations captured by our
model are both flexible and effective.
\item We performed a Bonferroni corrected paired t-test and show the
significance across all three levels for all four baselines.
Comparisons were relative to PACE, which has recently been adapted to
solve several different location-based recommendation problems.
\end{itemize}
\newcolumntype{P}[1]{>{\centering\arraybackslash}p{#1}}

\begin{table}[t]
	\footnotesize
	\centering
	\renewcommand{\arraystretch}{1.2}
	\mycaption{Impact of Parameters \adjustPara and $r_l$ on \textit{Beijing} dataset}
	\vspace{-2.5ex}
	\label{table-para-control}
	\begin{tabular}{P{0.4cm}P{1cm}P{0.5cm}P{0.5cm}P{0.5cm}P{0.5cm}P{0.5cm}P{0.5cm}}
		\toprule
		\multirow{2}{*}{Level} & \multirow{2}{*}{Metric} & \multicolumn{3}{c}{\adjustPara} & \multicolumn{3}{c}{$r_l$} \\
		\cmidrule{3-8}
		&                   & 0.6         & 1         & 1.4  & 50       & 150       & 250 \\ \midrule
		\multirow{2}{*}{$H_1$} & P@10   & 0.067  & 0.067  & \textbf{0.068} & 0.065 & 0.067 & \textbf{0.068} \\  
		& NDCG@10   & 0.161 & \textbf{0.162} & 0.162 & 0.153 & \textbf{0.162}  & 0.162\\ \midrule
		\multirow{2}{*}{$H_2$} & P@10   & 0.007 & \textbf{0.008} & 0.008 & 0.008 & \textbf{0.008} & 0.008\\  
		& NDCG@10   & 0.021  & \textbf{0.023}  & 0.023 & 0.021 & \textbf{0.023} & 0.023 \\ \midrule
		\multirow{2}{*}{$H_3$} & P@10   & 0.007 & \textbf{0.007} & 0.007 & 0.006 & \textbf{0.007} & 0.007  \\ 
		& NDCG@10   & 0.018 & \textbf{0.021} & 0.019 & 0.018 & \textbf{0.021} & 0.020\\ \bottomrule
	\end{tabular}
	\vspace{-4ex}
\end{table}

\vspace{-1ex}
\subsubsection{Tree Level Effects in \modelnameshort}
We also found that \modelnameshort performs relatively well at
both \poiTreeLayerOne and \poiTreeLayerTwo levels, since the 
upper level captures much richer information from the lower levels
using the attention mechanism.
Implicit feature representations of child POIs are aggregated from
child to parent, increasing the data available when learning the new
model.
In contrast, for POIs at the \poiTreeLayerThree level, these signals
are not available, and thus the overall performance compared with the
other baselines exhibits less dramatic performance improvements, but
is still effective.
\newcolumntype{P}[1]{>{\centering\arraybackslash}p{#1}}

\begin{table}[t]
	\footnotesize
	\renewcommand{\arraystretch}{1.2}
	\mycaption{Ablation study on the \textit{Beijing} dataset.}
	\vspace{-2.5ex}
	\label{table-exp3}
	\centering
	\begin{tabular}{ccccc}
		\toprule
		Level                  & Metric	& M1     & M2	& M3             \\ \midrule 
		\multirow{2}{*}{$H_1$} & P@10    	& 0.066    & 0.067	& 0.067  \\ 
		& NDCG@10 	& 0.156    & 0.160 & 0.162    \\ \midrule 
		\multirow{2}{*}{$H_2$} & P@10    & 0.007      & 0.008	& 0.008   \\  
		& NDCG@10    & 0.020        & 0.022  & 0.023 \\ \midrule 
		\multirow{2}{*}{$H_3$} & P@10    & 0.006      & 0.006	& 0.007  \\  
		& NDCG@10    & 0.010      & 0.011   & 0.021  \\ \bottomrule 
	\end{tabular}
	\vspace{-1em}
\end{table}

\vspace{-2ex}
\subsection{Hyper-parameter Studies (RQ2)} \label{subsec-paraTuning}
\subsubsection{Impact of Matrix Tradeoff Parameter}
Table~\ref{table-para-control} shows the results when varying
\adjustPara (in Eq.
\ref{equ-pred}) from $0.6$ to $1.4$, in order to control the tradeoff
between the feature-based check-in matrix and history-based check-in
matrix.
With the increase of \adjustPara, the effectiveness $NDCG@10$ of POI
recommendations at \poiTreeLayerTwo and \poiTreeLayerThree are more
sensitive than that at \poiTreeLayerOne.
From the results, we observe that the $NDCG@10$ at the
\poiTreeLayerThree level first goes up, and then begins to drop off.
Considering the holistic performance for all these three levels, our
model adopts the setting $\adjustPara=1$ that achieves its best
overall performance.

\vspace{-1ex}
\subsubsection{Impact of Embedding Size}
We also investigated the performance when varying the embedding size
$r_l$ from $50$ to $250$ in Table~\ref{table-para-control}.
The $NDCG@10$ of both \poiTreeLayerOne and
\poiTreeLayerTwo improved as expected since these levels have
access to additional information from the lower levels.
However, although the precision of \poiTreeLayerOne and
\poiTreeLayerTwo peak when $r_l=250$, the model training costs are
higher and may be more prone to overfitting.
In the remaining experiments, we chose $r_l=150$ since it offered the
best trade-off based on our internal experiments.

\vspace{-1ex}
\subsection{Recommendation Hints (RQ3)} \label{subsec-recExplain}
\begin{figure}[h]
	\vspace{-1.5em}
	\centering
	\subfloat[User-apsect hint]{
		\label{fig:explain:feature}
		\includegraphics[width=\heatmapscale\textwidth]{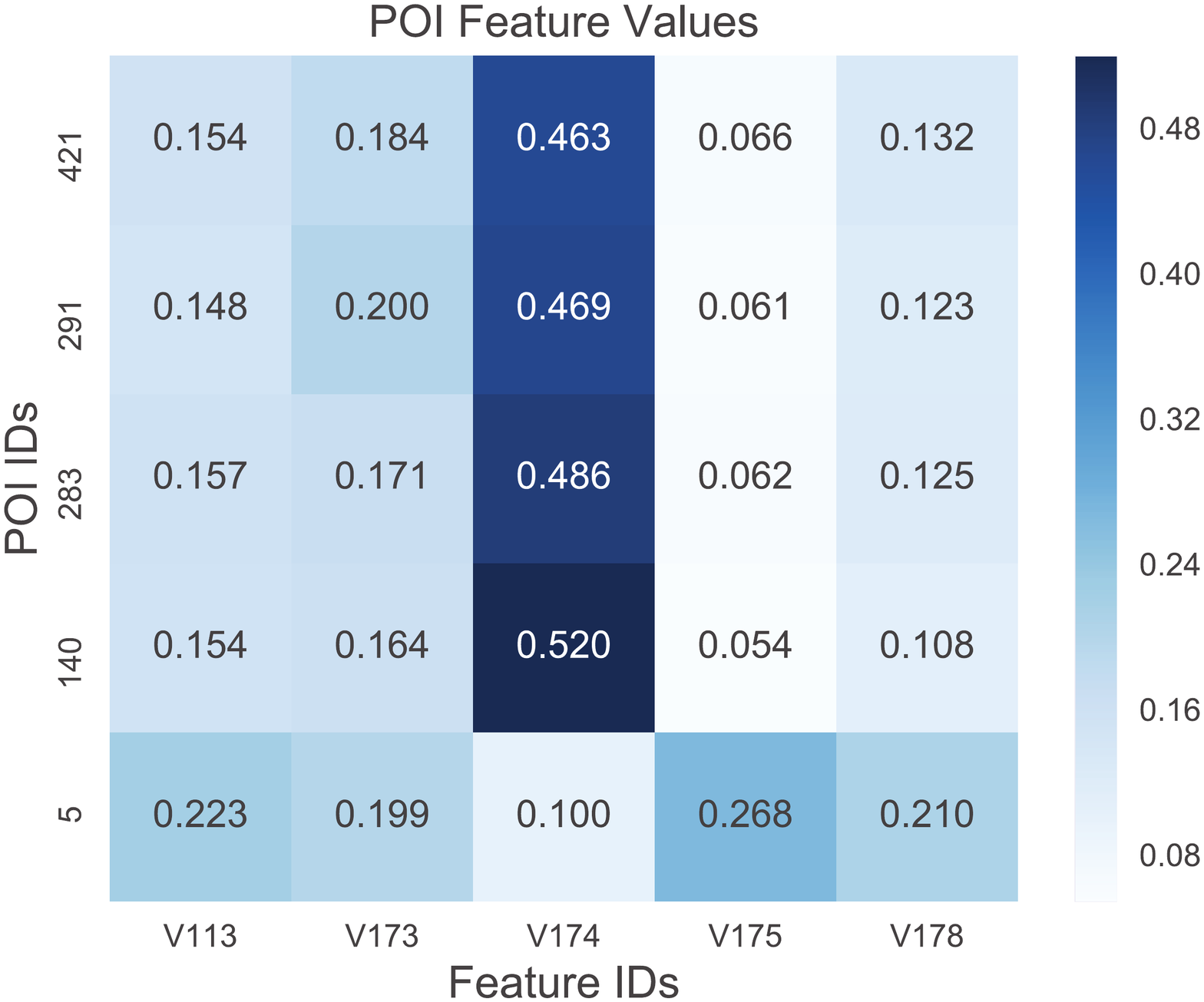}
	}
	\subfloat[POI-aspect hint]{
		\label{fig:explain:poi:att}
		\includegraphics[width=\heatmapscale\textwidth]{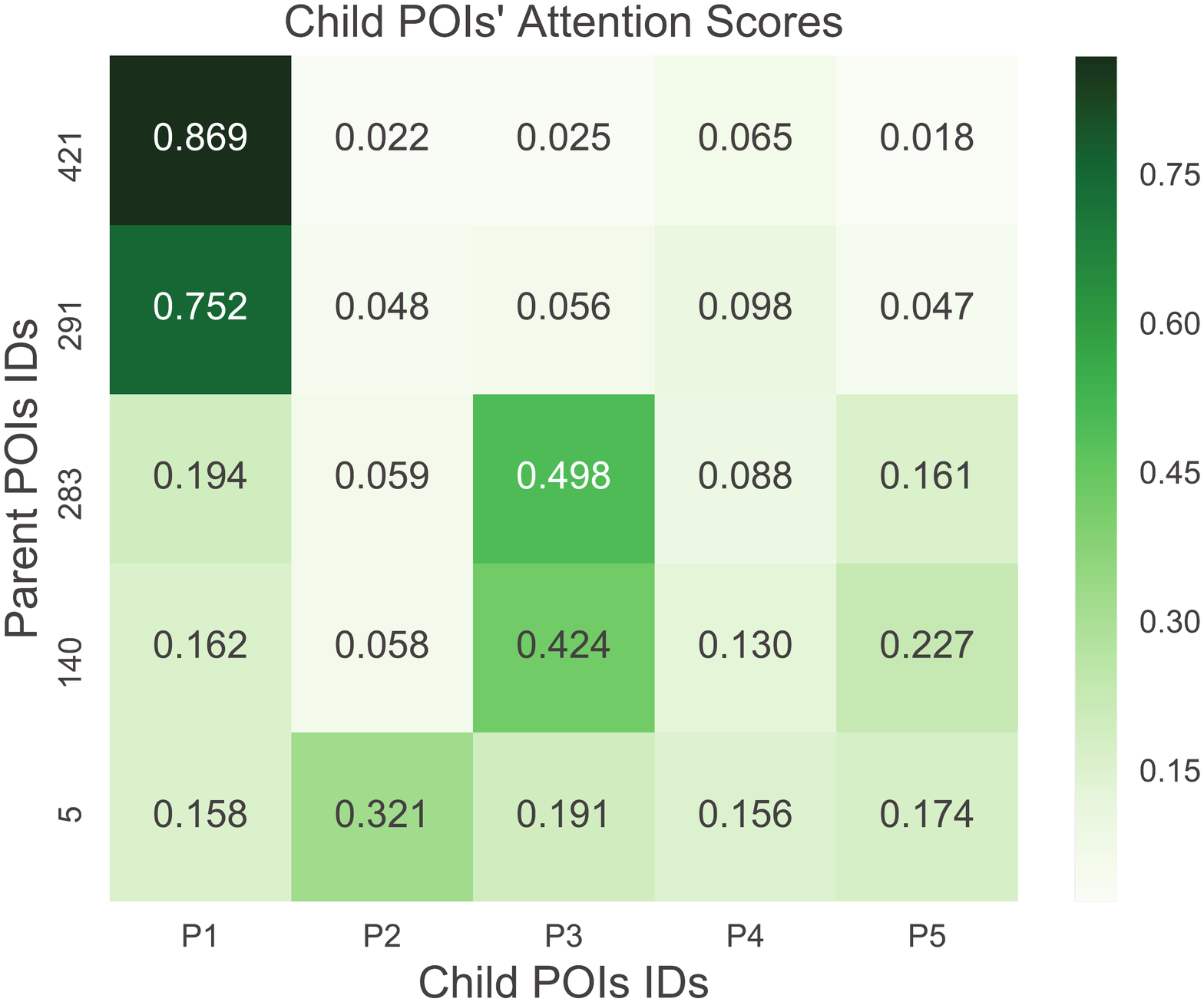}
	}
	\subfloat[Interaction-aspect hint]{
		\label{fig:explain:history}	
		\includegraphics[width=\heatmapscale\textwidth]{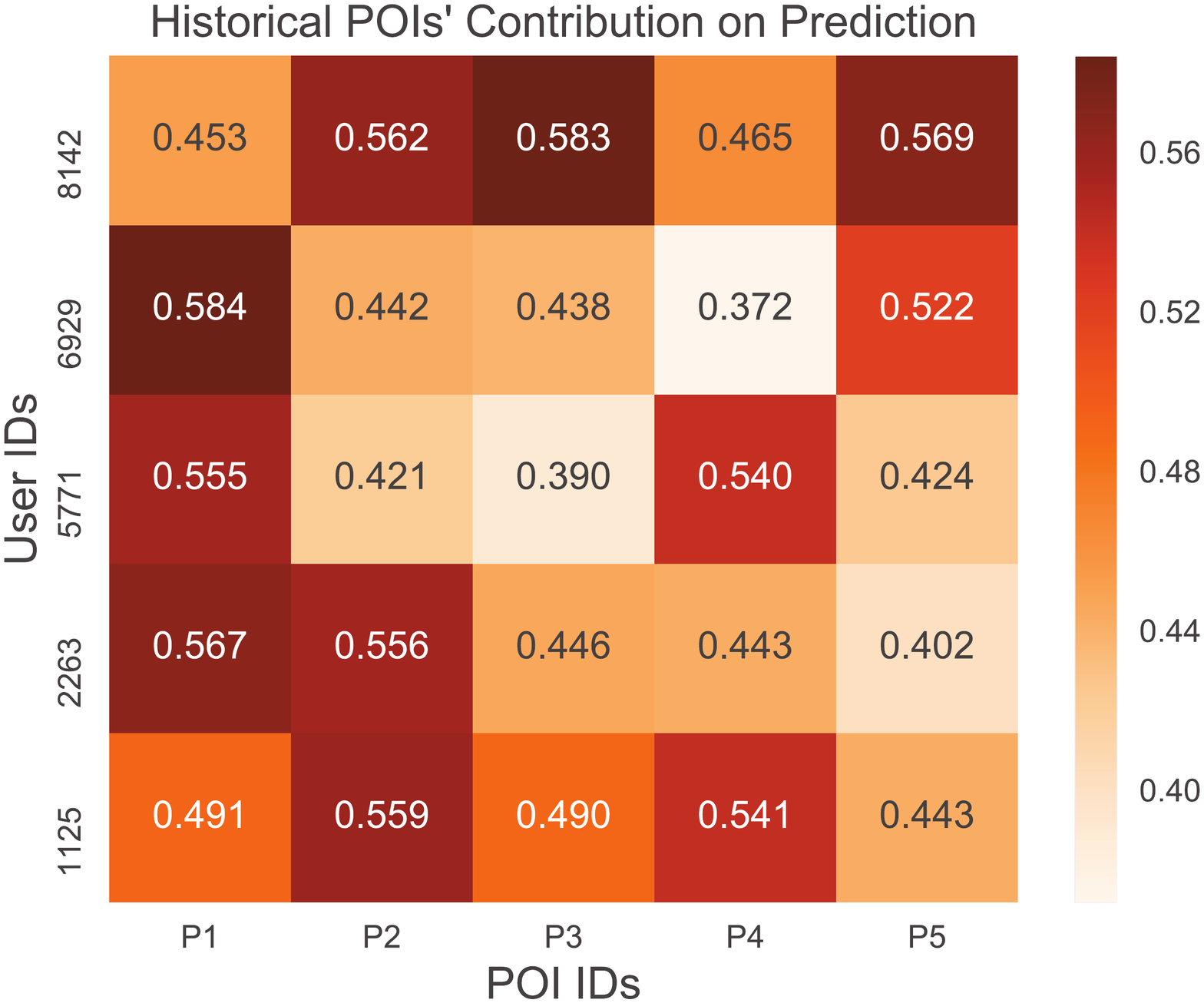}
	}
	\vspace{-2ex}
	\caption{Visualization heat maps of three recommendation hints on the \textit{Beijing} dataset. The larger a value is, the darker color its corresponding cell has. }
	\label{fig:explain}
	\vspace{-2em}
\end{figure}

We analyzed our model and created several heat maps to demonstrate
how recommendation hints might be created in Figure
\ref{fig:explain}.
All values are min-max normalized for direct comparisons in the
figure.


\vspace{-2.5ex}
\subsubsection{User-aspect hint}
Figure~\ref{fig:explain:feature} illustrates the POI feature
prediction values, where a row represents a recommended POI, and a
column denotes a POI feature.
In the figure, users were randomly sampled and we selected five user
features which best represented the sampled user preferences
according to the learned user feature prediction matrix $\vec{M}_u$.
Then we recorded the column numbers, which are V113, V173, V174,
V175, and V178.
We then recommended five POIs (i.e., 5, 140, 283, 291, and 421), and
extracted the POI feature prediction values from the learned POI
feature prediction matrix $\vec{M}_p^l$ by the corresponding recorded
columns accordingly (e.g., V113).
When examining the heatmap of the resulting POI feature values, we
can clearly observe the POI feature which has the highest value.
For example, when we recommended POI $421$st to the user, the V174
feature had the greatest contribution.

\subsubsection{POI-aspect hint}
Figure~\ref{fig:explain:poi:att} depicts the child POI attention
scores, where a row represents a recommended parent POI, and a column
denotes a child POI.
Specifically, we first chose the top-5 parent POIs recommended to a
user.
For each recommended parent POI, we analyzed the attention scores and
displayed the top-$5$ child POIs ($P1$-$P5$) that had the highest
attention score.
The score contribution ratios for each child POI are then displayed.
The child POI with the highest attention score can be interpreted as
follows.
When POI 421\textit{st} was recommended, we can observe that it had a
child POI $P1$ that was also important for that user.

\subsubsection{Interaction-aspect hint}
Figure~\ref{fig:explain:history} shows the contribution percentages
(i.e., $\expParaHis$) from the historical POIs used in the overall
prediction, where a row refers to a user, and a column is a
recommended POI.
In this experiment, we randomly chose five users.
For each user, we produce five recommended POIs ($P1$-$P5$).
If $\expParaHis$ w.r.t.\ a historical POI exceeds a fixed threshold
(say $0.5$), then we consider this historical POI to be a strong
influence on the final prediction.
Using the 1125\textit{th} user as a concrete example, the geospatial
influence from historical POIs had a strong influence on the
recommendations of $P2$ and $P4$.

\subsection{Ablation Study (RQ4)}
In this section, we present an ablation study to better understand
the influence of two core submodules: (i) child POI features
propagated to a parent POI bottom-up using the attention
mechanism (Section~\ref{subsubsec-feature_checkin}); (ii) the
geospatial influence factors between POIs derived from a POI context
graph, which map three different sources of spatial relationships
between any two POIs at the same POI level
(Section~\ref{subsubsec-history_checkin}).
We evaluated three variants of models with or without the above core
submodules: (1) M1: our model without both submodules (i) and (ii);
(2) M2: our model without submodule (ii); (3) M3: our model
\modelnameshort.
The experimental results when $k=10$ are shown in
Table~\ref{table-exp3}.
When comparing the model M1 with M2, we find that the attention
network mechanism indeed provides a substantial effectiveness
improvement in most cases.
Although \poiTreeLayerThree lacks the propagated child POI features,
the joint training across all POI levels still provides additional
performance benefits.
When comparing M2 and M3, we find that M3 also achieves consistent
performance improvements for $NDCG@10$, reaffirming the importance of
geospatial influence in the POI context graph.

\vspace{-2ex}
\section{Related Work} \label{sec:related}
POI recommendation has been intensively studied in recent years, with
a focus on how to integrate spatial and temporal properties
\cite{yuan2013time,yuan2014graph,yao2016poi}.
Recent advances in machine learning techniques have inspired
several innovative methods, such as sequential
embedding~\cite{zhao2017geo}, graph-based
embedding~\cite{xie2016learning}, autoencoder-based
models~\cite{ma2018point} and semi-supervised learning
methods~\cite{yang2017bridging}.
We refer the interested readers to a comprehensive survey
\cite{liu2017experimental} on POI recommendation.
In the remainder of this section, we review the most closely related
work to our own.

\myparagraph{Category-aware POI Recommendation}
Categories of POIs visited by a user often capture preferred
activities, thus they are important indicators to model user
preferences~\cite{zhou2019collaborative,xia2018intent,liu2019Hydra}.
\citet{liu2013personalized} exploited the transition patterns of
user preferences over location categories to enhance recommendation
performance.
Specifically, a POI category tree is built, where the top level has
\textit{food} or \textit{entertainment}, while the bottom level
includes \textit{Asian restaurant} or \textit{bar}.
\citet{zhao2017exploiting} showed that a POI has different
influences in different sub-categories.
Based on the hierarchical categories of each POI, they devised a
geographical matrix factorization method (which is a variant of
$\mathsf{GeoMF}$ \cite{lian2014geomf}) for recommendation.
The essential difference is that, each POI in
\cite{zhao2017exploiting} is still a single node but with multiple
influence areas for hierarchical categories, whereas in our problem a
POI has a tree structure constructed by spatial containment
relationship.
\citet{he2017category} adopted a two-step mode in their model, which
predicted the category preference of next POI first and then derived
the ranking list of POIs within the corresponding category.

However, these studies differ from our work.
They maintain a hierarchical structure of POI categories, but we
focus on how to exploit the spatial containment, rather than semantic
categories.

\myparagraph{Recommendation based on a Spatial Hierarchy}
The utility of exploiting hierarchical structures of either users or
items for item recommendation has been discussed in several prior
studies~\cite{lu2012exploiting,wang2015exploring,zhang2020semisupervised}.
Here we mainly highlight the key difference between existing
approaches involving spatial hierarchy and ours.

\citet{yin2017spatial} split the whole geographical area into a
spatial pyramid of varying grid cells at different levels.
The main purpose of such a spatial pyramid was to overcome the data
sparsity problem.
If the check-in data w.r.t. a region is sparse, then the check-in 
data generated by its ancestor regions can be used.
\citet{feng2017poi2vec} proposed a latent representation model to
incorporate geographical influence, where all POIs are divided into
different regions hierarchically and a binary tree is built over the
POIs in each region.
One major difference is that they aim to predict a set of users who
will visit a given POI in a given future period.
\citet{chang2018content} proposed a hierarchical POI embedding model
from two data layers (i.e., a check-in context layer and a text
content layer), neither of which is related with the tree structure
of POIs in our work.
\citet{zheng2011recommending} leveraged the hierarchy property of
geographic spaces to mine user similarity by exploring people's
movements on different scales of geographic spaces.
They assume that users who share similar location histories on
geographical spaces of finer granularities may be more correlated.
Therefore, these methods are not straightforward to cope with our
multi-level POI recommendation problem.

In summary, we are the first to define the multi-level POI
recommendation problem, and utilize a POI hierarchical tree structure
based on spatial containment to achieve POI recommendations from
varying spatial granularity.

\vspace{-1.5ex}
\section{Conclusion} \label{sec:conclusion}
In this work, we proposed and studied the \modelnamelong problem.
We show how to create POI recommendations at varying levels of
spatial granularity by constructing a POI tree, derived from various
spatial containment relationships between items.
Different from existing POI recommendation studies which support the
next-POI recommendation, we provide more recommendation strategies
which can be used directly by a wide variety of geographically based
recommendation engines.
To address this problem, we proposed a multi-task learning model
called \modelnameshort, where each task seamlessly combines two
subtasks: \taskOne \xspace and \taskTwo.
We also provide three different recommendation hint types which can
be produced using our model.
Finally, we compared our model with several state-of-the-art
approaches and two real-world datasets, thus demonstrating the
effectiveness of our new approach.
In future work, we will explore techniques to incorporate temporal
information into our model and further boost the effectiveness.

\vspace{-1ex}
\section{Acknowledgments}
This work was partially supported by NSFC 71531001 and 91646204, ARC DP180102050, DP200102611, DP190101113, and Google Faculty Research Awards. 
\vspace{-0.5em}

\bibliographystyle{ACM-Reference-Format}
\bibliography{ref_all}

\end{document}